\documentclass[numberedappendix,referee]{aastex}

\usepackage{graphicx}
\usepackage{natbib}
\usepackage{ulem}
\usepackage{txfonts}

\slugcomment{Submitted to the Astrophysical Journal}

\shorttitle{Small-scale emergence of magnetic flux into the quiet atmosphere}
\shortauthors{Mart\' inez Gonz\'alez \& Bellot Rubio}

\begin{document}

\title{Small magnetic loops connecting the quiet surface and the hot outer atmosphere of the Sun}

\author{M.\ J.\ Mart\' inez Gonz\'alez, R. Manso Sainz, A. Asensio Ramos}
\affil{Instituto de Astrof\'{\i}sica de Canarias, C/V\'{\i}a L\'actea s/n, 38200
La Laguna, Tenerife, Spain\\
Departamento de Astrof\'{\i}sica, Univ. de La Laguna, 38205, La Laguna, Tenerife, Spain}
\email{marian@iac.es}

\and

\author{L.\ R.\ Bellot Rubio}
\affil{Instituto de Astrof\' isica de Andaluc\' ia (CSIC), Apdo.\ 3004, 18080, Granada, Spain}

\begin{abstract}
Sunspots are the most spectacular manifestation of solar magnetism,
yet, 99\% of the solar surface remains 'quiet' at any time of the solar cycle.
The quiet sun is not void of magnetic fields, though; they are organized at  
smaller spatial scales and evolve relatively fast, which makes them difficult to detect.
Thus, although extensive quiet Sun magnetism would be a natural driver 
to a uniform, steady heating of the outer solar atmosphere, 
it is not clear what the physical processes involved 
would be due to lack of observational evidence.
We report the topology and dynamics of the magnetic field in very quiet
regions of the Sun from spectropolarimetric observations of
the Hinode satellite, showing a continuous injection of magnetic flux with a well
organized topology of $\Omega$-loop from below the solar surface into the upper layers. At first stages, when the loop travels across the photosphere, it has a flattened (staple-like) geometry and a mean velocity ascent of $\sim3$ km/s. When the loop crosses the minimum temperature region, the magnetic fields at the footpoints become almost vertical and the loop topology ressembles a potential field. The mean ascent velocity at chromospheric height is $\sim12$ km/s. The energy input rate of these small-scale loops in the lower boundary of the chromosphere is (at least) of $1.4\times 10^6-2.2\times 10^7$ erg cm$^{-2}$ s$^{-1}$. Our findings provide empirical evidence for solar magnetism as a multi-scale system, in which small-scale low-flux magnetism plays a crucial role, at least as important as active regions, coupling 
different layers of the solar atmosphere and being an important ingredient for chromospheric and coronal heating models.
\end{abstract}

% \section{}

Magnetic fields emerge onto the solar atmosphere as bipolar regions
of opposite polarities connected by field lines forming an $\Omega$-loop \citep{zwaan_85}. 
This emergence occurs in a broad range of spatial and magnetic flux scales,
ranging from the large active regions harboring sunspots, down to the lower scales of ephemeral regions. 
It seems natural to wonder if this pattern continues down to the resolution
limit and beyond, with smaller and weaker 
magnetic loops \citep{stenflo92}. 
However, the observation of relatively weak magnetic fields
at such small scales is extremely challenging, and 'quiet' Sun magnetism
is roughly characterized as disorganized or 'turbulent' \citep[e.g.][]{martin_88, stenflo92, javier_04}.
It has been argued that such turbulent fields could dissipate enough
magnetic energy to balance the radiative losses of the hot
chromosphere \citep{javier_04}, though no physical mechanism has been proposed.
Recently, evidence has been found that at least part of the magnetic fields
in the quiet Sun are well-organized as small bipoles on granular scales 
\citep{marian_07}, and that they are, in fact, highly dynamic \citep{marian_07, rebe_07, ishikawa_07, gomory_10}. This suggests the possibility of a truly multiscale solar magnetism with loop structures coupling the surface and atmosphere
at different spatial and time scales. Here, we show the detailed three-dimensional evolution
of the small-scale emerging magnetic loops in the quietest surface of the Sun and
reaching the outer atmosphere. Although the intermittent nature of these fields was argued to limit
their direct effect on the outer solar atmosphere \citep{schrijver_98}, we demonstrate that they may have important implications for the heating of the chromosphere.

The observations we analyse have been reported in \cite{marian_09}. They comprise 
full Stokes spectropolarimetry ($I$, $Q$, $U$, and $V$) in the Fe~{\sc i} \hbox{630~nm}
doublet, magnetograms in the \hbox{Mg~{\sc i}~b} 517.2~nm line, 
and filtergrams in the core of the \hbox{Ca~{\sc ii}~H} 396.9 nm line, 
all with instruments of the Solar Optical Telescope onboard the Hinode spacecraft. 
We use inversion techniques to derive the dynamics and topology of the observed small-scale 
magnetic flux emergence events. The magnetic field topology has been retrieved with the SIR inversion code \citep{basilio_92}. In each spectropolarimetric scan we choose the pixels with high enough polarimetric signal to get a reliable inversion. The inclinations at different pixels define a family of curves ('field lines') that are tangent to the inclinations at all pixels. Among all the possible field lines (the same one up to an arbitrary vertical shift), we choose the one passing through the 'height of formation' of the Fe~{\sc i} 630~nm line pair over the patch of solar surface where a significant amount of polarization is detected. When the linear polarization disappears we force the azimuth to lie on the line connecting both polarities and the inclination to vary linearly between the footpoints. Determining the magnetic field strength in low-flux areas from the inversion of noisy spectropolarimetric observations can be problematic due to degeneracies between thermodynamic and magnetic parameters \citep{marian_06}. In order to get a reliable estimate of the magnetic field strength 
we sacrifice spatial resolution to increase the signal-to-noise of the data.
We average the signal in all the pixels in a loop footpoint at different times and we have performed a Bayesian inversion based on a Milne-Eddington approach \citep{andres_marian_07}. 
However, the orientation of the magnetic field vector is very robust. 
In those frames in which linear (Stokes $Q$ and/or $U$) and circular polarisation 
(Stokes $V$) are detected, the magnetic field inclination and azimuth are
well determined. 

\begin{figure}[!ht]
\center
\includegraphics[width=8.9cm]{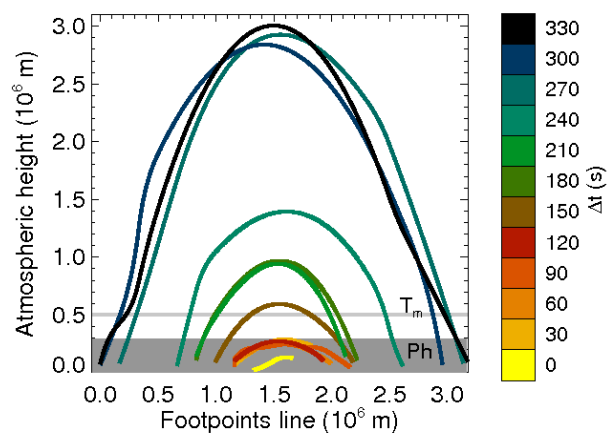}
\caption{Ascent of a small magnetic loop through the quiet solar atmosphere. Since the azimuth always lies along the line connecting the footpoints of the loop, we collapse the structure to the plane that containing this line. For each time the reconstructed loop is represented with different colors, from blue to orange as time increases. The grey area labeled as "Ph" represents the mean formation region of the 630 nm lines in the photosphere. The apex of the loop rises at a velocity of $\sim 3$ km s$^{-1}$ in the photosphere. The grey line labeled as "T$_\mathrm{m}$" represents the mean height for the temperature minimum region, associated with the formation of the \hbox{Mg\,I\,b} spectral line. Beyond the temperature minimum region, the apex of the loop ascends at a velocity of $\sim 12$ km s$^{-1}$, close to the sound speed at the chromosphere.}
\label{loop1}
\end{figure}

The reconstructed time evolution of such an emergence event is plotted in Fig. \ref{loop1}.
For simplicity, we show only the central field line of the whole bunch of lines. 
When the magnetic field emerges into the quiet surface the loop presents
a flattened, geometry and it maintains this geometry across the photosphere. 
This is qualitatively consistent with the topology observed in the discovery of low-lying loops \citep{marian_07}.
The mean ascent velocity within the photosphere is \hbox{$\sim3$ km s$^{-1}$}, 
while the footpoints separate between them at a velocity of 
6~km~s$^{-1}$. The loop has a magnetic field strength 
$\langle B\rangle=209\pm 7$~G occupying $30\pm 1$\% of the resolution element. 
Since such weak magnetic fields are subjected to convective motions, this velocity imbalance 
could explain the flattened topology of the magnetic loop while trying to travel across the high 
density photosphere. Beyond the photosphere the loop develops an arch-like geometry 
and its top rises at \hbox{$\sim 12$ km s$^{-1}$}, close to the 
sound speed in the chromosphere ($\sim 10$ km s$^{-1}$). 

The reconstruction satisfies various observational constraints. 
Linear polarization disappears at $\Delta t=120$~s, when the apex of the loop 
abandons the formation region of the 630~nm lines.
Two minutes later ($\Delta t=210-240$~s) some faint signals are observed in the
Mg~{\sc i}~b magnetogram at the same position of the footpoints in the photosphere.
Since longitudinal magnetograms are blind to horizontal fields,
we cannot determine when the apex reaches the temperature minimum.
However, at $\Delta t=180-240$~s, a larger portion of the reconstructed loop 
is predominantly vertical at the temperature minimum height. 
At that moment, the observed separation between footpoints in the Mg~{\sc i}~b magnetogram
is 1350~km while the distance between the loop footpoints is
between 1000 and 1730 km. 
At $\Delta t=750$~s, brightenings in the Ca~{\sc ii} H-line, corresponding to 
the low chromosphere, are identified co-spatially with the loop footpoints.
Although brightenings in this line are widely used as a proxy for magnetic activity,
we cannot associate their appearance with the loop reaching a given height.
If brightenings in \hbox{Ca\,II\,H} are produced only when the magnetic field is concentrated, 
it is possible that the loop could have reached the low chromosphere before $\Delta t= 750$ s but
we have been unable to detect it. The ascent to the corona remains unconstrained 
by our observations. 

\begin{figure*}[!ht]
\center
\includegraphics[width=12cm]{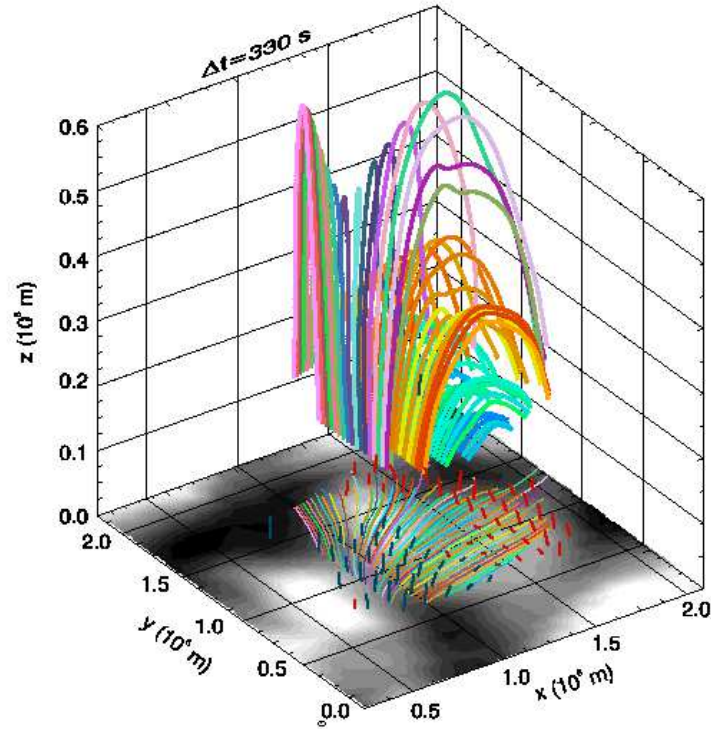}
\caption{Three-dimensional topology of the magnetic field over a granule. 
Continuum image at the bottom shows granular (bright) and intergranular (dark)
regions. Short lines indicate magnetic field orientation (blue for the footpoint
with positive, emergent polarity; red for the negative footpoint),
derived from inversion at the points with high enough spectropolarimetric signal. Representative field lines (tangent to these director vectors) are calculated
starting at the height of formation of the Fe~{\sc i} 630~nm lines at one
footpoint and followed until they reach the same height at the other end. 
Both footpoints happen to be connected. 
The projection of field lines on the solar surface appear as coloured lines
on the bottom plane, showing azimuth spreading over nearly 90$^\circ$.
From low-lying bluish lines to high-lying ones the magnetic field
fills most of the volume from the photosphere to the low-chromosphere. The colors of field lines have been used for the ease of eye.}
\label{loop2_1}
\end{figure*}

\begin{figure*}[!ht]
\center
\includegraphics[width=18.3cm]{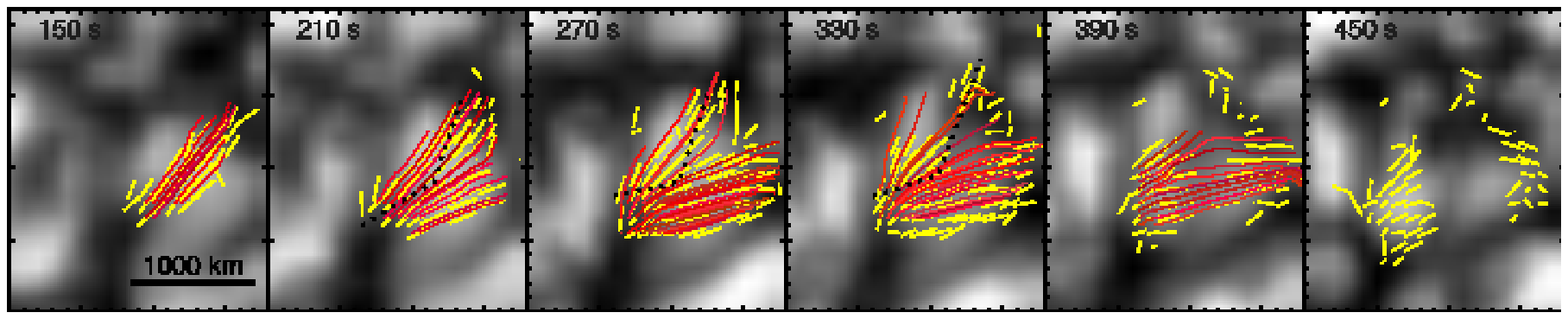}
\caption{Six snapshots of the evolution of a magnetic emergence event at granular scales. 
Yellow segments indicate the azimuth retrieved from inversions in
points with high enough spectropolarimetric signal.
Red lines show the projection of representative field lines.
During the first three minutes after emergence the magnetic field
maintains a characteristic $\Omega$-loop geometry with a well-defined
azimuth ($\sim 45^\circ$ in the first panel).
When reaching the intergranular lane, the upper-right footpoint
suffers strong shear while the other footpoint remains undivided;
consequently, field lines fan out, spreading in azimuth.
A dark lane appears at 210~s and fully develops beyond 270~s
(black dots), that splits the granule and the magnetic field
lines with it. Beyond 450~s it is not possible to follow the
three-dimensional reconstruction of the magnetic field lines
without further assumptions.
}
\label{loop2_2}
\end{figure*}

The dynamics of the emergence process can be more complicated.
The example shown in Figs. \ref{loop2_1} and \ref{loop2_2} is an extended event, 
rapidly expanding over the whole granule.
The flux emerges in a preexisting granule as a structure showing a simple bipolar loop with a clear preferred azimuth before developing a full
three-dimensional structure and dynamics (Fig. \ref{loop2_2}).
Then, one of the footpoints suffers strong shear when reaching the 
intergranular lane and the loop fans out, with the azimuth of individual field lines spanning 60$^\circ$.
At $\Delta t=330$~s the footpoints are detected in the \hbox{Mg~{\sc i}~b} magnetogram
coexisting with a (faint) linear polarization signal at the \hbox{630 nm} lines
from photospheric layers. This means that, although many field lines
have already reached the upper atmosphere, some of them are still in the
photosphere (Fig. \ref{loop2_1}). At $\Delta t=270$~s a dark lane appears across the granule and roughly along
the field lines.
The magnetic field vector has different azimuths at both sides of this
dark lane and as the plasma evolves, the granule splits in two and
so does the loop. 
At the end of the process the negative footpoint is divided
in two while the positive one still links both loops.
The azimuth of the magnetic field is parallel to the line
dividing both parts of the loop, showing that the evolution of the loop
is driven by the dynamics of the local granulation.
This is compatible with the relatively weak field strength 
$\langle B\rangle=176\pm 4$~G (filling factor $30\pm 1$\%)
found for this loop.

From Mg~{\sc i}~b magnetograms, the magnetic flux density for the loop of Fig. \ref{loop1} is 15~Mx~cm$^{-2}$. 
The temperature minimum marks just the lower boundary of the chromosphere, hence, this flux density 
can be used to compute the rate of magnetic energy injected to the chromosphere by the loop. 
Assuming the same magnetic filling factor as in the photosphere ($30$\%)
and an inclination of magnetic fields $~25^\circ$ as shown by the reconstructed
loops field lines at the temperature minimum region, the magnetic field strength is $B=55$~G. This number is consistent with estimating the magnetic field strength at the apex as $B'\approx BR^{2}/R'^2$ (from magnetic flux conservation), $R$ being the radius of the photospheric footpoints and $R'$ the radius of the loop cross section at the temperature minimum. From the reconstruction, we find that, around 500 km, $R^2/R'^2 \approx 0.25$. Therefore, the magnetic field strength must be 52 G. The magnetic field energy density associated to such a field is $E_{\rm mag}=B^2/8\pi\approx 120$~erg~cm$^{-3}$. The ascent velocity of the apex of the reconstructed loop at chromospheric heights is $v\sim 12$~km~s$^{-1}$, which gives a magnetic energy rate of $E_{\rm mag}v=1.4\times 10^8$~erg~cm$^{-2}$~s$^{-1}$ over the entire solar surface. However, we must correct this number by the portion of the area occupied by the emerging loops. The loop of Fig. \ref{loop1} is typical ---other events show similar magnetic fluxes, spatial and temporal scales. Thus, we take as the magnetic energy flux derived here as representative of the emergence process. From the analysis of the Hinode data used here, \cite{marian_09} report an emergence rate of 0.02~loops~h$^{-1}$~arcsec$^{-2}$, 23\% of them reaching the chromosphere. Assuming a mean life time of the loop in the chromosphere of $\sim 1050$~s, and estimating their area roughly from the mean separation of the footpoints ($\sim 1800$~km), then just a fraction $\alpha\sim$1\% of the solar surface undergoes an emergence event at any given time, which leads to a magnetic energy injection of $\alpha E_{\rm mag}v\approx 1.4\times 10^6$~erg~cm$^{-2}$~s$^{-1}$. For comparison, this is one order of magnitude short than the $10^7$~erg~cm$^{-2}$~s$^{-1}$ radiative losses estimated for the chromosphere from non-local thermodynamic equilibrium semi-empirical models of the solar atmosphere \citep{anderson_89}.

Our estimate of the magnetic energy injection is rather conservative and it will probably
increase with more sensitive observations. The absolute number of detected events will increase 
with higher signal-to-noise observations or if we observe at more sensitive spectral 
regions as the infrared. In fact, \cite{marian_07} report 0.0098 loops arcsec$^{-2}$ from spectropolarimetric observations of the Fe~{\sc i} \hbox{1.56 $\mu$m} lines without temporal resolution. Assuming that both infrared and visible loops are of the same nature and last $~2$~minutes as a recognisable full loop in the photosphere, the estimated emergence rate in the infrared is 0.3 loops h$^{-1}$ arcsec$^{-2}$. The magnetic energy injection is then $2.2\times 10^7$~erg~cm$^{-2}$~s$^{-1}$, which is of the same order of the estimated radiative losses for the chromosphere.

The role of the very quiet Sun in heating the outer atmosphere comprises three physical aspects: the generation and transport of mechanical energy from the photosphere to the chromosphere, and the dissipation of this energy in those layers. We have shown that the flux emerges as a three-dimensional myriad of field lines that, despite their weak field strengths and hence their distortion by plasma motions, adopt the form of $\Omega$-loops and reach the chromosphere. The energy injection of these loops in the chromosphere is at least of the same order of magnitude as the radiative losses in that layer. This magnetic energy must be transformed into heat by a physical mechanism. Recent magneto-hydrodynamical (MHD) simulations suggest that the small loops reach chromospheric heights and get reconnected with the local expanding magnetic fields, heating the plasma and generating MHD waves that propagate into the corona \citep{isobe_08}. Another source of heating could be the dissipation of magnetic energy at electric current density \citep{sami_03} and should be quantified in future works. We can estimate that, at photospheric layers, using Amp\`ere's Law, the mean value of current sheets is of about \hbox{15-45 mA m$^{-2}$}.

It has been suggested that the radiative losses are significantly larger in the chromosphere than in the corona \citep{ishikawa+tsuneta2009}, thus the chromospheric heating represents a bigger challenge
as it requires a large energy input. Furthermore, it seems that the process(es) responsible for the
coronal heating must also act in the chromosphere \citep{pontieu_09}. Taken together, these
arguments suggest that small-scale loops emerging in the quiet Sun may have important consequences not only
for the heating of the chromosphere, but also that of the corona.

\begin{acknowledgements}

We acknowledge financial support from the Spanish MICINN through the project AYA2007-63881. 
We are grateful to all the observers who participated in the {\it Hinode} Operation Plan 14, 
both at ISAS/JAXA and at the ground-based telescopes. Hinode is a Japanese mission developed and launched 
by ISAS/JAXA, with NAOJ as a domestic partner, and NASA and STFC (UK) 
as international partners. It is operated by these agencies in cooperation with ESA and NSC (Norway). 

\end{acknowledgements}

% \bibliographystyle{apj}
% \bibliography{/home/marian/latex/mybib}

\end{document}